# View about consumption tax and grandchildren

Eiji YAMAMURA (Seinan Gakuin University, JAPAN)

Postal Address: Department of Economics, Seinan Gakuin University, 6-2-92 Sawaraku Nishijin,

Fukuoka 814-8511, Japan.

Telephone number: 81-92-823-4543

Fax number: 81-92-823-2506

e-mail address: yamaei@seinan-gu.ac.jp

Abstract

In Japan, the increase in the consumption tax rate, a measure of balanced public finance, reduces the inequality of fiscal burden between the present and future generations. This study estimates the effect of grandchildren on an older person's view of consumption tax, using independently collected data. The results show that having grandchildren is positively associated with supporting an increase in consumption tax. Further, this association is observed strongly between granddaughters and grandparents. However, the association between grandsons and grandparents depends on the sub-sample. This implies that people of the old generation are likely to accept the tax burden to reduce the burden on their grandchildren, especially granddaughters. In other words, grandparents show intergenerational altruism.

1. Introduction

Many researchers argue that a consumption-based tax system is preferable to an income-based one (Aaron and Gale, 1996; Gravelle, 1991; Hall and Rabushka, 1995; Summers, 1981). In an aging society, consumption tax is regarded as financing government expenditures to alleviate intergenerational inequality because the ratio of older people to the total population has rapidly increased (Watanabe et al., 2015). In most developed countries where demographic and fiscal problems are critical, consumption-based tax plays a significant role in realizing sustainable economies.

For instance, in a rapidly aging society like Japan, public debt and fiscal burden have remarkably increased to finance social security expenditure (Kitao and Mikoshiba, 2020). The fiscal balance of the Japanese government has rapidly deteriorated, and a growing fiscal deficit could become a burden on future generations. Accordingly, it is necessary to set a consumption tax rate of around 30%–45% to achieve fiscal sustainability (Braun and Joines, 2015). In 2030, the government of Japan could reduce the budget deficit by approximately half of the consumption tax rate rises from 10% to 15% (İmrohoroğlu et al., 2019). According to Kitao (2011), an increase in consumption taxes transfers wealth from old to young consumers. Consequently, future generations will experience welfare gains, and present generations, particularly older consumers, experience welfare losses[1]. The increase in consumption tax rates can be considered an intergenerational redistribution policy.

We confront the conflict of interest between the present and future generations,

---

[1] Future generations will be better off with pension reform, and delaying reforms will reduce welfare of the future generations in Japan (Kitao, 2017).

analogous to the conflict between high-income and low-income groups. There is a trade-off by increasing the consumption taxes between grandparents and their grandchildren if they have selfish motives. Young grandchildren cannot display their intentions and views because of their immaturity. Inevitably, grandchildren would encounter more difficult situations with larger fiscal burdens than their grandparents. There is a possibility of strategic behavior in parent–child relations (Horioka et al., 2018). For instance, parents behave according to their expectations of their child's caregiving in the future. Such strategic decision-making is unlikely in the grandparent-grandchild relationship. Therefore, grandparents' views about consumption taxes are unlikely to be selfish concerning the influence of their grandchildren. However, subjective views and perceptions are significantly affected by culture and traditional societal values (Alesina et al., 2004; Luttmer and Singhal, 2011). For instance, older people may have altruistic or dynastic motivations to redistribute wealth to their children through bequest behavior (Horioka, 2000, 2002, 2014). Existing studies indicate that fathers with daughters favor females (Oswald and Powdthavee, 2010; Washington, 2008) and promote a sustainable society (Cronqvist and Yu, 2017). This indicates the child-to-parent behavioral influence may lead parents to be altruistic toward their children[2].

In an aging society, older persons are expected to play a significant role in the workplace[3]. The role played by grandparents within a family has also increased (e.g., He et al., 2018; Mutchler et al., 2007; Zeng and Xie, 2014), such as caregiving for grandchildren (e.g., Boca et al., 2018; Feng and Zhang, 2018; Greenfield, 2011)[4].

---

[2] Mothers influence sons' views and preferences about working women and gender-equalization (Fernandez et al., 2004; Kawaguchi and Miyazaki, 2009).
[3] In the situation of a rapid decline in the labor force caused by the unprecedented speed of demographic aging, old persons are required in the labor supply (Kitao and Mikoshiba, 2020).
[4] Families anticipate grandmothers to care for their grandchildren to promote the labor participation of mothers. To examine this, many studies analyzed the relationship between grandmothers'

Inevitably, grandparents and grandchildren spend time together forming intimate relations. Hence, grandchild-to-grandparent behavioral influence is also critical in achieving a sustainable society that benefits grandchildren and future generations. It is valuable to analyze whether the grandchild-to-grandparent relationship promotes intergenerational redistribution policy, such as increased consumption tax in the long-term. Hence, this study examines the effect of grandchildren on their grandparents' view of increasing consumption tax in Japan.

Based on independently collected individual-level data, in addition to basic demographic and economic information, a subjective view about consumption tax policy and family structure is collected. The key findings indicate that older persons with grandchildren are more likely to support an increase in consumption tax than those who do not. Further, this tendency is far stronger for older persons with granddaughters than those with grandsons. This implies that altruistic grandparents are motivated to redistribute wealth to the future generation because the probability that granddaughters may earn more than grandsons is low. Many empirical studies have explored how and why people prefer income redistribution (e.g., Corneo and Gruüner, 2002; Luttmer and Singhal, 2011; Ravallian and Lokshin, 2000; Yamamura, 2012). Preference for intergenerational redistribution has not been sufficiently explored in empirical analysis thus far. The study's contribution is to identify the reasons for preference for intergenerational redistribution and altruism stemming from having grandchildren.

The remainder of this paper is organized as follows. Section 2 proposes the study's

---

childcare and the labor supply of mothers (e.g., Aparicio-Fenoll and Vidal-Fernandez, 2015; Garcia-Moran and Kuehn, 2017; Ho, 2015; Posadas and Vidal-Fernández, 2012; Rupert and Zanella, 2018). Moreover, the grandmothers' caregiving has influenced grandparents' health status (Di Gessa et al., 2016a; 2016b; Ku et al., 2012; Reinkowski, 2013), mortality (Christiansen, 2014), participation in social activities (Arpino and Bordone, 2017) and cognitive functioning (Ahn and Choi, 2019).

hypotheses. Section 3 describes the data and presents the empirical methodology. The estimation results and their interpretations are presented in Section 4. The final section offers some reflections and conclusions.

## 2. Testable Hypotheses

Due to the unprecedented speed of aging in society, grandparents are expected to care for grandchildren (Boca et al., 2018; Feng and Zhang, 2018; Greenfield, 2011). Consequently, grandparents have opportunities to participate in physical exercise and mental activity, improving their subjective well-being (Coall and Hertwig, 2011; Dunifon et al., 2020; Powdthavee, 2011; Silverstein et al., 2003; Wang et al., 2019)[5]. Therefore, through caregiving, grandparents develop an intimate relationship with their grandchildren.

Piketty (1995) theoretically indicates that expectations of upward and downward mobility determine individual attitudes toward redistribution. Bénabou and Ok (2001) proposed the "prospect of upward mobility (POUM)" hypothesis that people who expect to move up the income scale will not favor a distributive policy even if they are currently poor. Ravallion and Lokshin (2000) found that people expecting their welfare to fall in the future tend to support redistribution, even if they are currently wealthy. Recent studies by Alesina et al. (2018) bridged the intergenerational social mobility and redistribution preferences. They used an experimental method to provide evidence that respondents support more redistribution after seeing pessimistic information about their mobility. This tendency is observed only for left-wing respondents and not for right-wing ones.

---

[5] Brunello and Rocco (2019) found that informal care of grandchildren reduced grandparents' subjective well-being.

In Japan, public debt, and fiscal imbalance have rapidly increased, leading to an increased fiscal burden for the future generation. Consumption tax is a widely acknowledged method and an effective measure to redistribute from the present generation to the future generations, reducing intergenerational inequality (Kitao, 2011; Kitao and Mikoshiba, 2020; Watanabe et al., 2015). Similar to bequest behaviors (Horioka, 2000, 2002, 2014, 2019), grandparents prefer redistribution to their grandchildren.

Therefore, we propose *Hypothesis 1*.

*Hypothesis 1: Having a grandchild leads old persons to support an increase in the consumption tax rate.*

Daughters increase parents' life satisfaction and form more intimate parent–child relations than sons (Pushkar et al., 2014). Gender differences in children significantly influenced parents' views to favor females. This plausibly holds for the grandparent–grandchild relationship. The Global Gender Gap Index 2020 rankings indicate that Japan was ranked 121st among 153 countries when gender equality was evaluated from various indices (World Economic Forum, 2020)[6]. According to the Basic Survey on Wage Structure[7], the female wage level was 74 % of the male wage level in 2019. Females could realize upward mobility if they married high-income persons. However, according to the Japan National Census, the percentage of unmarried women increased consistently from

---

[6] Even for two-income households in Japan, mothers worked from home to take care of their children (primary school), whereas husbands did not change their lifestyle during the unexpected school closure under the COVID-19 pandemic(Yamamura and Tsustsui, 2021).
[7] https://www.jil.go.jp/kokunai/statistics/timeseries/html/g0406.html (accessed on January 28, 2021)

20% in 1970 to 61% in 2015 for the 25–29 age group, and from 10% in 1970 to 35% in the 30–34 age group. Considering gender inequality and the high rate of unmarried women, grandparents predict that their granddaughters would face economic difficulty compared to their grandsons. This leads grandparents to redistribute their current wealth to future generations.

Hence, *Hypothesis 2* is raised.

*Hypothesis 2: Having a granddaughter is more likely to lead old persons to support an increase in the consumption tax rate than those with a grandson.*

Apart from the expected economic conditions grandchildren will face in the future, cross-gender effects are observed as daughters cause fathers to prefer policies favoring females (Oswald and Powdthavee, 2010; Washington, 2008) and to invest in sustaining society (Cronqvist and Yu, 2017). This applies to the grandfather–granddaughter relationship. Thus, *Hypothesis 3* is raised.

*Hypothesis 3: The effect of granddaughters on grandfathers supporting consumption tax is larger than that on grandmother.*

In the United States, the relationship between the income of grandparents and grandchildren depends on gender (Olivetti et al., 2018). The authors found that the social and economic status of grandsons is strongly influenced by that of paternal grandfathers than maternal grandfathers. This behavior is consistent if parents' bequests are determined by dynastic motivation (Horioka, 2000, 2002). That is, paternal grandparents are more likely to favor grandsons than granddaughters.

Thus, *Hypothesis 4* is proposed.

*Hypothesis 4: Paternal grandparents have a dynastic motivation to support consumption tax if they have grandsons.*

### 3. Data

3.1. Data collection

The data were collected independently, at an individual level, through an internet survey in July 2016. The Nikkei Research Company was commissioned, owing to their reputation among Japanese researchers and experience with academic surveys, to survey a representative sample of Japanese people aged 18 to 68.[8] The survey was kept open to collect at least 10,000 observations. Eventually, 12,176 observations were obtained. In the original survey, the sample's demographic composition covered 18–67 years of age. However, this study explores how grandparents' views about consumption tax differ from those of old persons who were not grandparents. Therefore, we used sub-samples of old persons who may have grandchildren.

Figure 1(1) illustrates the distribution of ages using a sample where respondents have children and grandchildren. Those over 60 years of age occupy approximately 70%. Further, 95% of respondents who have grandchildren are at least over 50 years of age. All respondents equal to or below 40 years of age did not have grandchildren. Therefore, we limit the sample to those over 40 years when we conduct the estimations. Figure 1(2) illustrates the rates of the age of respondents who have children but do not have any grandchildren and indicates a declining rate as respondents age. Younger respondents and respondents without children are unlikely to have grandchildren. In this study, the sample is limited to parents who are over 60 years because their child is possibly a parent. That is, there is a possibility that the respondents have a grandchild. Further, we limited the sub-sample of those who answered questions related to this study.

---

[8] A 2015 government survey on the use information technology indicates that over 90% of working-age Japanese wee internet users. Accordingly, the bias due to the exclusion of non-users is unlikely to be large. See the Statistics Bureau, Ministry of Internal Affairs and Communications http://www.soumu.go.jp/johotsusintokei/statistics/statistics05.html (access on April 5, 2018).

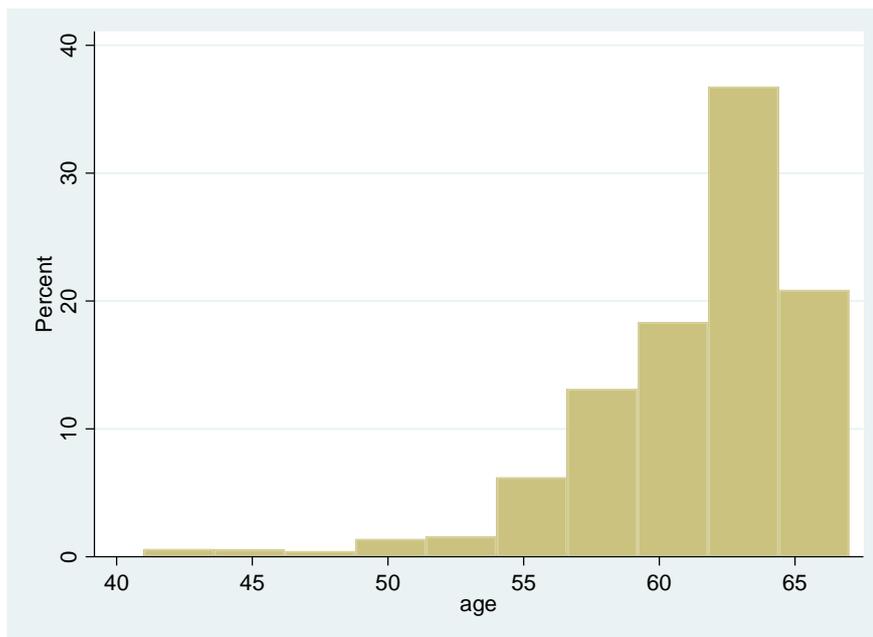

Figure 1(1). Distribution of ages of respondents with children and one grandchild at least. (Respondents over 40 years old)

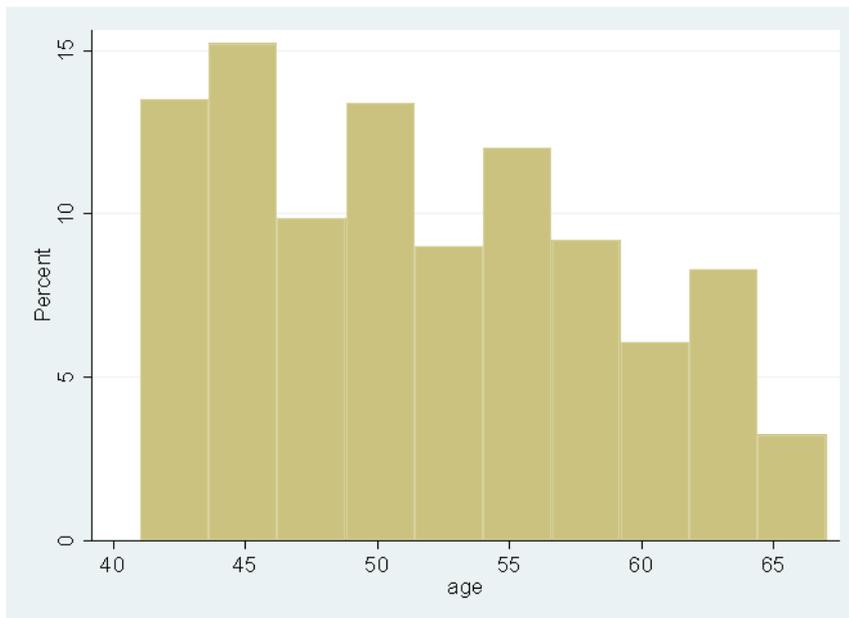

Figure 1(2). distribution of ages when respondents have their child but do not have grandchildren. (Respondents over 40 years old)

Observations were reduced to 4,125, 2,684, and 1,102 if respondents had children and were over 40, 50, and 60 years, respectively. Unfortunately, we do not have data on the age of the grandchildren. However, we can predict the ages of respondents' grandchildren based on official data (Ministry of Health, Labour and Welfare, 1980, 2017). Figures 1(1) and 1(2) indicate that the ages of respondents who have grandchildren are approximately 63. They were about 25 years of age in 1980. In 1980, the average age of women having their first child was approximately 26. Therefore, their first child was 38 years old when the surveys were conducted. In 2016, the average age of women with their first child was approximately 30. Based on this sub-sample, Figure 2 shows the distribution of the number of grandchildren. Approximately two-thirds of the respondents had grandchildren. The sample was further limited to respondents who were over 60 years of age, with children over 30 years, ensuring respondents are grandparents and making them comparable to others. Then, the observations were condensed to 623. Additionally, various sub-samples were used for a more detailed examination.

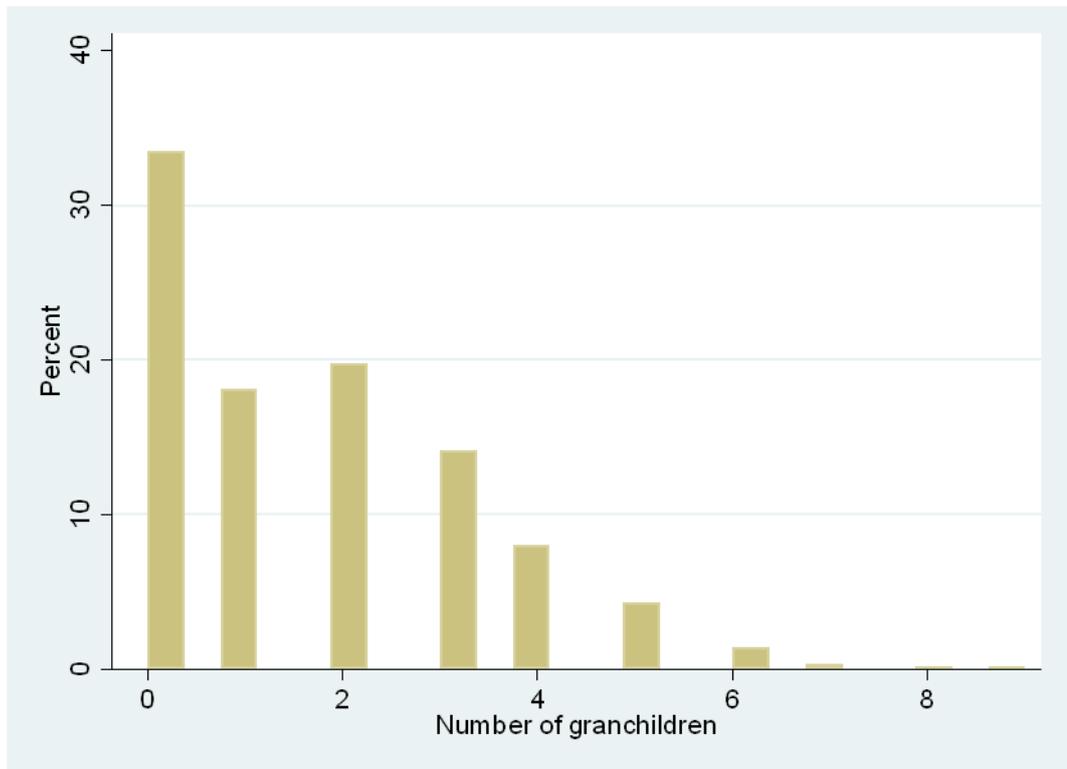

Figure 2. Distribution of the number of grandchildren.

Note: The sample is limited to respondents over 60 years of age and with children over 30 years.

3.2. Data definition and descriptive statistics

The questionnaire includes various questions to collect basic variables such as respondents' ages, genders, job status, marital status, educational background, household income, and residential prefectures. Additionally, information about family members such as the number of children and grandchildren, their gender, and age of the youngest child were collected.

It is widely acknowledged that an increase in consumption tax would necessitate fiscal sustainability (Braun and Joines, 2015; İmrohoroğlu et al., 2016; İmrohoroğlu et al.,

2019). This study examines a subjective view of the argument on consumption tax. Hence, the question about the key variable is:

*"To what degree do you agree with the statement that an increase in consumption tax cannot be avoidable?"*

Preferences for consumption tax (CON TAX) were measured by the degree of support ranging from 1 (strongly disagree) to 5 (strongly agree).

Based on a subsample of respondents who were over 60 years of age and had children, Figure 3 shows the comparison of the distribution of "Preferences for consumption tax" between those who had grandchildren and others. As shown in Figure 3, respondents who have grandchildren are more likely to choose larger values than their counterparts. This is consistent with *Hypothesis 1*. However, Table 1 only compares the mean difference but does not consider various other factors. For closer examination, the study controls for various variables using regression estimations in Section 4.

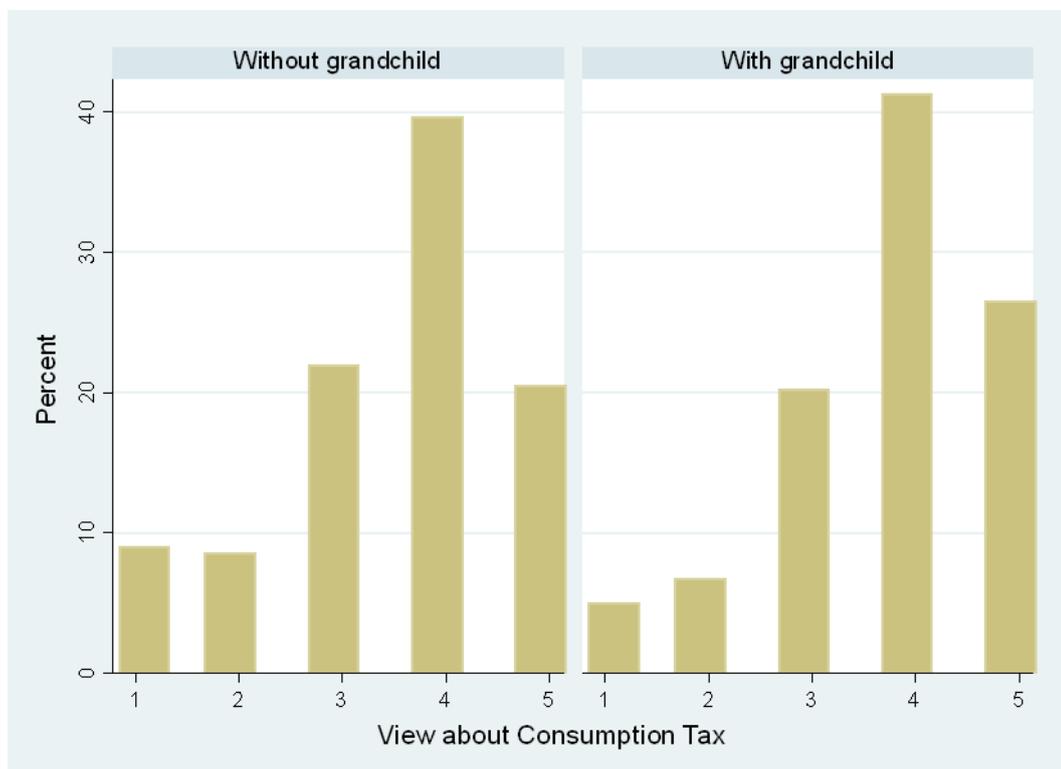

Figure 3. Distribution of view about the consumption tax. Comparison between those with grandchildren or not.

Note: The sample is limited to respondents over 60 years of age and with children over 30 years.

Characteristics may differ according to the number of grandchildren or genders of grandchildren. Using the same subsample used in Figures 2 and 3, Table 1 indicates the balance check. Panel A suggests no difference in respondents' ages, the youngest child's age, household income, and schooling years. Deleting respondents having both sons and daughters from the sample in Panel A, Panel B compares variables between respondents with only sons and only daughters. This also does not show a difference in the variables between the two groups.

3.3. Empirical specifications

Our baseline model assesses how the presence of grandchildren is related to grandparents' preference of consumption tax, testing *Hypothesis 1*. The estimated function takes the form:

$$\text{CON TAX}_i = \alpha_0 + \alpha_1 \text{ GRAND CHILDREN}_i + X_i B + u_i.$$

The dependent variable, CON TAX, is a proxy for the preference of consumption tax. The suffix *i* denotes the individuals. $X_i$ denotes the control variable, and *B* denotes the vector of the estimated coefficients. The control variables are the individuals' age (AGE) and their square (AGESQR), respondents' gender dummy (MALE), dummies for schooling years (EDU), 12 job status dummies, 17 household income dummies, and marital status dummies.

The key independent variable was GRAND CHILDREN, indicating the number of grandchildren. From *Hypothesis 1,* the sign of the coefficient is predicted to be positive. The effect of the grandchildren's genders is compared to test *Hypothesis 2*. In the alternative specification, GRAND CHILDREN is replaced by the number of grandsons (GRAND SONS) and granddaughters (GRAND DAUGHTERS). Having children depends on parents' decisions, not on grandparents. Therefore, GRAND CHILDREN is considered an exogenous variable. However, there is a possibility that the gender of the child reflects parents' and grandparents' preferences if they have dynastic motivation. In this case, parents continue to be motivated to have children until they have a son. Consequently, the number of children may have endogenous biases. Hence, instead of the number of grandchildren, grandsons, and granddaughters, a dummy is incorporated (GRAND CHILDREN DUMMY, GRAND SONS DUMMY, and GRAND DAUGHTERS DUMMY).

*Hypothesis 3* is tested by including the cross-terms GRAND DAUGHTERS*MALE. The expected sign was positive. *Hypothesis 4* was tested using a sub-sample composed of respondents with only sons over 30 years. Based on the sample, the expected sign of the GRAND SONS DUMMY was positive.

4. **Estimation Results**

The baseline specification results using the subsample of those over 40, 50, and 60 years of age are presented in Tables 2 and 3. Table 2 shows key variables related to the number of grandchildren and basic control variables. Tables 3–8 indicate only the results of key variables and the grandchildren dummy. Table 1 shows that the GRAND CHILDREN coefficient yields a positive sign while being statistically significant at the

1% level, in columns (1)–(3). Therefore, the number of grandchildren is positively associated with an increase in the consumption tax. After dividing grandchildren into granddaughters and grandsons, GRAND DAUGHTERS produces a significant positive sign, columns (4)–(6), while GRAND SONS yields a significant positive sign only in column (6). As for the control variables, only INCOM consistently shows a positive sign and is statistically significant at the 1% level in all columns. This reflects "the fact that consumption taxes are more regressive than progressive income taxes. In particular, fundamental tax reform may have unfavorable effects on low-income households, who spend more of their resources on consumption' (Kitao, 2011, p. 63).

As seen in Table 3, rather than the number of grandchildren, the existence of grandchildren is analyzed. As in Table 2, the GRAND CHILDREN DUMMY shows a significant positive sign in all estimations. A significant positive sign is observed only for GRAND DAUGHTERS DUMMY, but not for GRAND SONS DUMMY. Considering the results in Tables 2 and 3, *Hypotheses 1* and *2* are supported.

To test *Hypothesis 3*, cross-terms between the granddaughters dummy and male dummy variables, GRAND DAUGHTERS DUMMY * MALE and grandsons dummy and male dummy variables, GRAND SONS DUMMY * MALE, are added to the specification in Table 3. GRAND DAUGHTERS DUMMY * MALE shows a significant positive sign in column (2), although not in columns (1) and (3). This is, to a certain extent, consistent with *Hypothesis 3*. However, the statistical significance depends on the sub-sample, and thus the results are not robust.

For thorough testing of *Hypotheses 1* and *2,* further estimations were conducted. The sub-sample of respondents over 60 years of age was further limited to those with children over 30 years by using the information on the youngest child's age. Additionally, only those respondents who had only one child or were childless were included because the genders of the first child were considered more exogenous. However, information only about the numbers of children, sons, and daughters were available. Therefore, the gender of the children cannot be identified if there are multiple children. For instance, the gender of the first child is unknown if the respondents have one son and one daughter. Therefore, respondents having multiple children in the sub-sample are removed to consider the first child's gender. The results in Table 5 show a positive sign for all variables. Furthermore, statistical significance was observed in GRAND CHILDREN, GRAND DAUGHTERS, GRAND CHILDREN DUMMY, and GRAND DAUGHTERS DUMMY in all columns. These results are similar to those in Tables 2 and 3. The CON TAX is a discrete and ordered variable. In this case, the ordered probit model is more appropriate than the OLS model. Using the same specification and sub-samples, ordered probit estimations are conducted for the robustness check. In Panel A of Tables 6 and 7, the sign and statistical significance of the key variables are the same as those in Table 5. Therefore, *Hypotheses 1* and *2* are strongly supported by the results of Tables 2–3 and 5–7. However, the marginal effects, shown in Panel B of Tables 6 and 7, indicate the effects of key variables more closely than OLS results in Table 5.

In columns (1) and (3) of Panel B in Table 7, the probability of "CON TAX=5," that older persons with grandchildren strongly agreeing with the increase in consumption tax, is approximately 8–9% higher than those who do not have grandchildren. Meanwhile, the probability of "CON TAX=1," that older persons with grandchildren strongly disagreeing with the increase in consumption tax, is 4% lower than those who do not have grandchildren. Absolute values of marginal effects for the probability of other choices are 0.02 to 0.04, meaning that having a grandchild leads to a 2–4 % difference in the decision.

Therefore, the effect of having a grandchild on the probability of choosing to strongly agree is approximately three times larger than other probabilities. The results of having a granddaughter are similar to those of having a grandchild. The effect of having a granddaughter on the probability of choosing to strongly agree is two to three times larger than other probabilities. However, having a grandson does not influence the probability of choosing a view about consumption tax at all.

As a whole, grandparents are unlikely to have altruistic motivation to support consumption tax when they have a grandson. However, as Horioka (2000, 2002, 2014, 2019) tested, grandparents may have dynastic motivations to redistribute wealth to grandsons. That is, *Hypothesis 4* is tested by conducting estimations focusing on whether respondents' sons have a child or not. Table 8 indicates the results using various subsamples where respondents with daughters are not included. Except for column (1), GRAND DAUGHTERS DUMMY shows a significant positive sign, as observed in other tables. Interestingly, GRANDSONS DUMMY shows a positive sign and statistical significance, seen in columns (3) and (4), where we limit the sample to respondents with children or not. This means that grandparents support the consumption tax increase to favor their son's son, explained by the Dynastic motivation. Therefore, *Hypothesis 4* is supported. The marginal effects shown in Panel B suggest that the marginal effect of having a son is sizable and almost equivalent to that of having a daughter. The probability of "CON TAX=5," that older persons with only sons and grandsons strongly agree with the increase in consumption tax, is approximately 16% higher than those with sons but no grandsons. Notably, the results depend on whether multiple grandsons are included in the sample. Insignificant results of GRAND SONS in columns (1) and (2) suffer from endogenous biases.

Tables 2–8 show that grandparents' altruism toward their grandchildren is for consumption tax policy. However, cross-gender effects between grandfathers and granddaughters and dynastic motivation are partially observed but are not robust. The findings of this study are inconsistent with the argument that the Japanese are less altruistic than other countries (Horioka, 2000, 2002, 2014, 2019). However, this study does not compare the findings in Japan with those in other countries. For closer examination, data from other developed countries should be used for a comparative study.

These findings have the following policy implications. In a society with fewer children, the rate of old persons having grandchildren decreases while the population of old persons increases. Therefore, old persons are less likely to support the policy to realize a sustainable society because they are less likely to have altruistic motivation toward grandchildren. The existence of grandchildren leads grandparents to imagine the future where their descendants live, and thus, they behave altruistically. In this regard, increasing the birth rate is critical to sustain society and look toward a bright future.

5. **Conclusions**

In an aging society with increasing public debt and fiscal burden, an increase in consumption tax rates is required to reduce the intergenerational gap to maintain economies in Japan. However, older persons in the present generation would not support an increase in consumption tax if they were rational and selfish, hampering long-term sustainable economies. How can we deal with this problem?

Consumption tax is considered a measure for achieving intergenerational redistribution from present to future generations. It seems plausible that people of the current generation are motivated to support the consumption-tax-based system depending on the probability of the existence of their descendants. Contrary to parent–child relations, grandparent–grandchild relations are unlikely to be related to strategic decision making based on, for instance, parents' expectations about their child's caregiving in the future. Therefore, the influence of having grandchildren on their grandparents' views about consumption taxes is based on altruism rather than selfish motivation. The influence of grandchildren on their grandparents' preferences for consumption taxes is examined using independently collected data.

The major findings are: as a whole, having a grandchild is positively associated with grandparents' support for increased consumption taxes. This implies that grandparents show intergenerational altruism. Further, this association is observed only between granddaughters and grandparents, but not between grandsons and grandparents. This can be interpreted as grandparents' significant motivation to reduce their granddaughter's burden because granddaughters are less likely to earn more than grandsons in the future due to the large gender gap in Japan. This is consistent with the argument that the expectation of moving down the income scale will favor a distributive policy (Bénabou and Ok, 2001; Piketty, 1995).

The effects of grandfather–granddaughter relations and dynastic motivation are partially observed but are not robust. The effects of using experimental or quasi-experimental analyses need to be scrutinized. This study assumes that grandparents expect granddaughters to earn lesser than grandsons when they become adults. Compared with other developed countries, there are larger gender gaps in labor participation in Japan and other facets such as political participation (World Economic Forum, 2020). For a closer examination, it is necessary to conduct estimations using a dataset covering countries where gender gaps are smaller. Further, an extended analysis should be conducted covering cultural and social norms from the viewpoint of comparative analysis, as done by Alesina et al. (2018) and Horioka (2000, 2014). These issues remain to be addressed in future works.

## 6. Acknowledgment

We would like to thank an anonymous reviewer for the insightful comments that provoked a thorough revision and Editage (www.editage.jp) for English language editing.

## 7. Funding

This study was supported by a Grant-in-Aid for Scientific Research B (Grant No. 16H03628) from the Japan Society for the Promotion of Science.

## 8. Declaration of Interest

Table 1. Balance check

Panel A.

| Number of grandchildren | Ages | The Youngest Child's ages | Income | Schooling years |
|---|---|---|---|---|
| 0 | 63.4 | 35.1 | 4.66 | 14.3 |
| 1 | 63.5 | 34.9 | 4.65 | 14.2 |
| 2 | 63.6 | 36.0 | 4.09 | 13.9 |
| 3 | 63.5 | 35.9 | 4.17 | 14.0 |
| 4 | 63.2 | 34.9 | 4.22 | 14.0 |
| 5 | 63.7 | 36.2 | 3.90 | 13.4 |
| 6 | 63.8 | 34.6 | 4.22 | 13.6 |
| 7 | 64.0 | 36.0 | 3.50 | 13.0 |
| 8 | 63.0 | 35.0 | 3.00 | 12.0 |
| 9 | 66.0 | 32.0 | 6.00 | 16.0 |
| P-value (F-test) | 0.66 (0.73) | 0.68 (0.75) | 0.36(1.1) | 0.18 (1.40) |

Panel B. Sample is limited to those who have only grandsons or granddaughters (deleting both genders of grandchildren)

| Genders of grandchildren | Ages | Child's ages | Income | Schooling years |
|---|---|---|---|---|
| Grand-sons | 63.4 | 35.1 | 4.66 | 14.3 |
| Grand-daughters | 63.5 | 34.9 | 4.65 | 14.2 |
| t-value (F-test) | 0.66 (0.73) | 0.68 (0.75) | 0.36(1.1) | 0.18 (1.40) |

Table 2. Results of baseline specification. Sub-sample of respondents with children. Test for altruistic motivation (OLS model)

|  | (1) | (2) | (3) | (4) | (5) | (6) |
|---|---|---|---|---|---|---|
|  | AGE>=40 | AGE>=50 | AGE>=60 | AGE>=40 | AGE>=50 | AGE>=60 |
| GRAND CHILDREN | 0.05*** (3.20) | 0.05*** (2.88) | 0.08*** (3.65) |  |  |  |
| GRAND DAUGHTERS |  |  |  | 0.07** (2.08) | 0.66* (1.98) | 0.09*** (2.99) |
| GRAND SONS |  |  |  | 0.04 (1.22) | 0.03 (1.11) | 0.07** (2.21) |
| AGE | −0.02 (−0.45) | −0.08 (−0.48) | −0.61 (−0.37) | −0.02 (−0.47) | −0.09 (−0.50) | −0.61 (−0.37) |
| AGE SQR | 0.0003 (0.60) | 0.0008 (0.52) | 0.004 (0.37) | 0.0003 (0.61) | 0.0008 (0.54) | 0.004 (0.37) |
| INCOM | 0.08*** (7.15) | 0.07*** (5.43) | 0.07*** (2.71) | 0.08*** (7.14) | 0.07*** (5.42) | 0.07*** (2.71) |
| EDU | 0.03** (2.41) | 0.01 (1.08) | 0.01 (0.86) | 0.03** (2.41) | 0.01 (1.08) | 0.01 (0.86) |
| MALE | 0.02 (0.35) | 0.10 (1.60) | 0.22** (2.04) | 0.02 (0.34) | 0.10 (1.60) | 0.22** (2.04) |
| Observations | 4,125 | 2,684 | 1,102 | 4,125 | 2,684 | 1,102 |
| R-square | 0.05 | 0.04 | 0.05 | 0.05 | 0.04 | 0.05 |

Note: Numbers in parentheses are t-values calculated using robust standard errors clustered in residential prefectures. **, and *** indicate significance at the 5%, and 1% levels, respectively. Various control variables are included, such as the respondent's marital status and job status dummies. However, these results have not been reported.

Table 3. Results using dummy variable of grandchildren, instead of the number of grandchildren. Sub-sample of respondents with children. Test for altruistic motivation (OLS model)

|  | (1) | (2) | (3) | (4) | (5) | (6) |
|---|---|---|---|---|---|---|
|  | AGE>=40 | AGE>=50 | AGE>=60 | AGE>=40 | AGE>=50 | AGE>=60 |
| GRAND CHILDREN DUMMY | 0.10* (1.97) | 0.10* (1.72) | 0.25*** (3.41) |  |  |  |
| GRAND DAUGHTERS DUMMY |  |  |  | 0.12* (1.79) | 0.12* (1.70) | 0.17** (2.55) |
| GRAND SONS DUMMY |  |  |  | −0.001 (−0.02) | −0.007 (−0.12) | 0.09 (1.37) |
| Observations | 4,125 | 2,684 | 1,102 | 4,125 | 2,684 | 1,102 |
| R-square | 0.05 | 0.04 | 0.05 | 0.05 | 0.04 | 0.05 |

Note: Numbers in parentheses are t-values calculated using robust standard errors clustered in residential prefectures. **, and *** indicate significance at the 5%, and 1% levels, respectively. The set of independent variables is equivalent to that in Table 2. However, these results have not been reported.

Table 4. Results of cross-terms. Sub-sample of respondents with children. Test for cross-gender effects (OLS model)

|  | (1) AGE>=40 | (2) AGE>=50 | (3) AGE>=60 |
| --- | --- | --- | --- |
| **GRAND DAUGHTERS DUMMY * MALE** | 0.20 (1.62) | 0.22* (1.86) | 0.14 (1.04) |
| **GRAND SONS DUMMY * MALE** | 0.05 (0.50) | 0.03 (0.24) | 0.04 (0.26) |
| Observations | 4,125 | 2,684 | 1,102 |
| R-square | 0.05 | 0.04 | 0.05 |

Note: Numbers in parentheses are t-values calculated using robust standard errors clustered in residential prefectures. **, and *** indicate significance at the 5%, and 1% levels, respectively. The set of independent variables is equivalent to that in Table 2. However, these results have not been reported.

Table 5. Results using sub-sample of respondents with children who were over 30 years old. Test for altruistic motivation (OLS model)

| | (1) Children over 30 ages AGE>=60 | (2) Children over 30 ages AGE>=60 | (3) Children over 30 ages AGE>=60 | (4) Children over 30 ages AGE>=60 | (5) Children over 30 ages AGE>=60 GRND CHILDREN=1 or 0 | (6) Children over 30 ages AGE>=60 GRND CHILDREN=1 or 0 |
|---|---|---|---|---|---|---|
| GRAND CHILDREN | 0.08*** (3.31) | | | | | |
| GRAND DAUGHTERS | | 0.11** (2.61) | | | | |
| GRAND SONS | | 0.05 (1.25) | | | | |
| GRAND CHILDREN DUMMY | | | 0.29*** (3.28) | | 0.28** (2.33) | |
| GRAND DAUGHTERS DUMMY | | | | 0.20* (1.98) | | 0.38** (2.08) |
| GRAND SONS DUMMY | | | | 0.07 (0.62) | | 0.19 (1.35) |
| Observations | 623 | 623 | 623 | 623 | 322 | 322 |
| R-square | 0.07 | 0.07 | 0.07 | 0.06 | 0.11 | 0.11 |

Note: Numbers in parentheses are t-values calculated using robust standard errors clustered in residential prefectures. **, and *** indicate significance at the 5%, and 1% levels, respectively. The set of independent variables is equivalent to that in Table 2. However, these results have not been reported.

Table 6. Results of the number of grandchildren using sub-sample of respondents who were over 60 years old. Further, they have children who were over 30 years old. Test for altruistic motivation (Ordered Probit model)

Panel A.

|  | (1) Children over 30 ages AGE>=60 | (2) Children over 30 ages AGE>=60 |
|---|---|---|
| **GRAND CHILDREN** | 0.08*** (3.06) |  |
| **GRAND DAUGHTERS** |  | 0.11*** (2.62) |
| **GRAND SONS** |  | 0.05 (1.29) |
| Observations | 623 | 623 |
| Pseudo R-square | 0.05 | 0.03 |

Panel B. Marginal effects

|  | (1) | (2) |
|---|---|---|
| **GRAND CHILDREN** (CON TAX=1) | −0.01*** (−2.86) |  |
| (CON TAX=2) | −0.01*** (−3.13) |  |
| (CON TAX=3) | −0.01*** (−3.01) |  |
| (CON TAX=4) | 0.004** (2.32) |  |
| (CON TAX=5) | 0.02*** (3.11) |  |
| **GRAND DAUGHTERS** (CON TAX=1) |  | −0.01** (−2.31) |
| (CON TAX=2) |  | −0.01*** (−2.82) |
| (CON TAX=3) |  | −0.02*** (−2.77) |

|  |  |
|---|---|
| (CON TAX=4) | 0.006** (1.97) |
| (CON TAX=5) | 0.03*** (2.73) |
| **GRAND SONS** (CON TAX=1) | −0.01 (−1.32) |
| (CON TAX=2) | −0.004 (−1.26) |
| (CON TAX=3) | −0.007 (−1.26) |
| (CON TAX=4) | 0.003 (1.27) |
| (CON TAX=5) | 0.01 (1.28) |

Note: Numbers in parentheses are z-values calculated using robust standard errors clustered in residential prefectures. **, and *** indicate significance at the 5%, and 1% levels, respectively. The set of independent variables is equivalent to that in Table 2. However, the results have not been reported.

Table 7. Results of grandchildren dummies using sub-sample of respondents over 60 years old. Further, they have children who were over 30 years old. Test for altruistic motivation. (Ordered Probit model)

Panel A.

|  | (1) Children over 30 ages AGE>=60 | (2) Children over 30 ages AGE>=60 | (3) Children over 30 ages AGE>=60 GRND CHILDREN=1 or 0 | (4) Children over 30 ages AGE>=60 GRND CHILDREN=1 or 0 |
|---|---|---|---|---|
| GRAND CHILDREN DUMMY | 0.30*** (3.44) |  | 0.28*** (2.61) |  |
| GRAND DAUGHTERS DUMMY |  | 0.19** (2.02) |  | 0.39** (2.09) |
| GRAND SONS DUMMY |  | 0.08 (0.75) |  | 0.18 (1.53) |
| Observations | 623 | 623 | 322 | 322 |
| Pseudo R-square | 0.03 | 0.03 | 0.05 | 0.05 |

Panel B.

|  | (1) | (2) | (3) | (4) |
|---|---|---|---|---|
| GRAND CHILDREN DUMMY (CON TAX=1) | −0.04*** (−2.97) |  | −0.04** (−2.19) |  |
| (CON TAX=2) | −0.03*** (−3.57) |  | −0.03** (−2.51) |  |
| (CON TAX=3) | −0.04*** (−3.56) |  | −0.04*** (−2.80) |  |
| (CON TAX=4) | 0.02** (2.43) |  | 0.02** (2.05) |  |
| (CON TAX=5) | 0.09*** (3.49) |  | 0.08*** (2.62) |  |
| GRAND DAUGHTERS DUMMY (CON TAX=1) |  | −0.02* (−1.80) |  | −0.05* (−1.95) |

| | | |
|---|---|---|
| (CON TAX=2) | −0.02** (−2.15) | −0.04** (−2.02) |
| (CON TAX=3) | −0.03*** (−2.18) | −0.05** (−2.17) |
| (CON TAX=4) | 0.01 (1.60) | 0.03* (1.80) |
| (CON TAX=5) | 0.06** (2.11) | 0.11** (2.13) |
| GRAND SONS DUMMY (CON TAX=1) | −0.01 (−0.75) | −0.02 (−1.37) |
| (CON TAX=2) | −0.01 (−0.75) | −0.02 (−1.52) |
| (CON TAX=3) | −0.01 (−0.74) | −0.02 (−1.57) |
| (CON TAX=4) | 0.004 (0.77) | 0.01 (1.33) |
| (CON TAX=5) | 0.02 (0.74) | 0.05 (1.52) |

Note: Numbers in parentheses are z-values calculated using robust standard errors clustered in residential prefectures. **, and *** indicate significance at the 5%, and 1% levels, respectively. The set of independent variables is equivalent to that in Table 2. However, these results have not been reported.

Table 8. Results of grandchildren dummies using sub-sample of respondents with sons who were over 30 years old but do not have daughters. Test for dynastic motivation. (Ordered Probit model)

Panel A.

|  | (1) Sons over 30 ages. | (2) Sons over 30 ages AGE>=60 | (3) Sons over 30 ages. GRNDCHILD=1 or 0 | (4) Sons over 30 ages. GRNDCHILD=1 or 0 AGE>=60 |
|---|---|---|---|---|
| GRAND DAUGHTERS DUMMY | 0.29 (1.58) | 0.33* (1.82) | 0.69** (2.20) | 0.86** (2.35) |
| GRAND SONS DUMMY | 0.14 (0.85) | 0.01 (0.08) | 0.65** (2.42) | 0.63** (2.55) |
| Observations | 237 | 197 | 151 | 120 |
| Pseudo R-square | 0.07 | 0.08 | 0.10 | 0.10 |

Panel B.

|  | (1) | (2) | (3) | (4) |
|---|---|---|---|---|
| GRAND DAUGHTERS DUMMY (CON TAX=1) | −0.04 (−1.47) | −0.05 (−1.60) | −0.08* (−1.88) | −0.11* (−2.00) |
| (CON TAX=2) | −0.03 (−1.51) | −0.03* (−1.75) | −0.06** (−1.96) | −0.06** (−1.94) |
| (CON TAX=3) | −0.04* (−1.71) | −0.04** (−2.04) | −0.09** (−2.31) | −0.11*** (−2.61) |
| (CON TAX=4) | 0.02 (1.52) | 0.02* (1.64) | 0.06* (1.88) | 0.05 (1.56) |
| (CON TAX=5) | 0.08 (0.59) | 0.09} (1.83) | 0.17** (2.19) | 0.22** (2.31) |
| GRAND SONS DUMMY (CON TAX=1) | −0.02 (−0.89) | −0.001 (−0.08) | −0.07** (−2.02) | −0.08** (−2.27) |
| (CON TAX=2) | −0.01 (−0.82) | −0.001 (−0.08) | −0.05** (−1.98) | −0.04* (−1.82) |
| (CON TAX=3) | −0.02 (−0.81) | −0.002 (−0.08) | −0.09*** (−2.78) | −0.08*** (−2.98) |
| (CON TAX=4) | 0.01 (0.79) | 0.01 (0.08) | 0.05** (2.05) | 0.04* (1.69) |
| (CON TAX=5) | 0.04 (0.86) | 0.04 (0.08) | 0.16** (2.41) | 0.16** (2.45) |

Note: Numbers in parentheses are z-values calculated using robust standard errors clustered in
residential prefectures. **, and *** indicate significance at the 5%, and 1% levels, respectively. The set of independent variables is equivalent to that in Table 2. However, these results have not been reported.